\documentclass[Afour,times]{sagej}
\usepackage{moreverb,url}
\usepackage[colorlinks,bookmarksopen,bookmarksnumbered,citecolor=red,urlcolor=red]{hyperref}
\usepackage{makecell, multirow, subcaption, caption}

\newcommand\BibTeX{{\rmfamily B\kern-.05em \textsc{i\kern-.025em b}\kern-.08em
T\kern-.1667em\lower.7ex\hbox{E}\kern-.125emX}}

\def\volumeyear{20251}

\begin{document}

\runninghead{Zhang and Zhao, et al.}

% \title{Simulation in Conversational Recommender Systems: A systematic Review}
\title{A Literature Review on Simulation in Conversational Recommender Systems}

\author{Haoran Zhang\affilnum{1}, Xin Zhao\affilnum{2}, Jinze Chen\affilnum{3}, and Junpeng Guo\affilnum{1}}

\affiliation{\affilnum{1}Tianjin University\\
\affilnum{2}University of Electronic Science and Technology of China\\
\affilnum{3}University of Washington}

\corrauth{Junpeng Guo, Tianjin University, Tianjin, China.}

\email{guojp@tju.edu.cn}

\begin{abstract}
Conversational Recommender Systems (CRSs) have garnered attention as a novel approach to delivering personalized recommendations through multi-turn dialogues. This review developed a taxonomy framework to systematically categorize relevant publications into four groups: dataset construction, algorithm design, system evaluation, and empirical studies, providing a comprehensive analysis of simulation methods in CRSs research. Our analysis reveals that simulation methods play a key role in tackling CRSs’ main challenges. For example, LLM-based simulation methods have been used to create conversational recommendation data, enhance CRSs algorithms, and evaluate CRSs. Despite several challenges, such as dataset bias, the limited output flexibility of LLM-based simulations, and the gap between text semantic space and behavioral semantics, persist due to the complexity in Human-Computer Interaction (HCI) of CRSs, simulation methods hold significant potential for advancing CRS research. This review offers a thorough summary of the current research landscape in this domain and identifies promising directions for future inquiry.
\end{abstract}


\maketitle

\section{Introduction}

CRSs are recommendation task-oriented information access systems that primarily interact with users through dialogue. Unlike conventional Recommender Systems (RSs), which may rely on static user data and profiles, CRSs can extract users' precise and fine-grained preferences through multi-turn conversations. Furthermore, CRSs differ from conventional dialogue systems (such as chit-chat systems) in their focus on delivering personalized recommendations. As depicted in Figure 1a, research on CRSs has garnered substantial attention in recent years, largely due to advancements in natural language processing and the emergence of pre-trained language models.

CRSs possess the following characteristics: (1) System Architecture. As depicted in Figure 1b, current CRSs architectures are primarily categorized into two types: pipeline and end-to-end models. Despite the elegance of end-to-end CRSs, the isolation between dialogue and recommendation tasks complicates the construction of a unified end-to-end model. We do not rule out the possibility of future research to build a more effective unified model, but most current CRSs are pipeline-based. (2) Complex Interaction. HCI in CRSs inherits all forms and properties of conventional RSs and dialogue systems, including conversations, clicks, multi-turn interactions, etc., thereby increasing complexity. (3) Vague Ground Truth. In CRSs, clear ground truth is almost non-existent. Systems must optimize within users' near-infinite dialogue content and preference states \cite{yoon-etal-2024-evaluating}. These characteristics present significant research challenges for CRSs in data, algorithms, evaluation, and empirical research. Simulation methods, including data and user simulations, offer effective solutions to these challenges. In addition, several recent studies have begun to leverage the capabilities of large language models (LLMs) to develop CRS simulations \cite{yoon-etal-2024-evaluating, wang-etal-2023-rethinking-evaluation, liang-etal-2024-llm}. Given this emerging trend, we argue that summarizing the current research progress in this area is timely and essential.

A series of CRSs-focused reviews are related to our work, but most were completed before the remarkable progress of pre-trained language models \cite{GAO2021100, Jannach2023, PRAMOD2022117539}, so they missed the phase of rapid development of CRSs driven by LLMs. The remaining relevant studies either centred on CRSs evaluation but ignored simulation methods \cite{Jannach-chen-2022}, or covered user simulation for evaluating information access systems but couldn't focus on our topic \cite{Balog-zhai-f18}. In short, no literature has deeply and systematically summarized simulation methods and their applications in CRSs, to our knowledge. Thus, our Literature Review aims to address two key issues:

\textbf{(1) What is the role of simulation methods in CRSs research?}

\textbf{(2) What are the challenges and opportunities in existing studies?}

% \begin{enumerate}
% \item[(1)]

% \item[(2)]

% \item[(3)]
% \end{enumerate}

% \begin{figure*}[htbp]
% \centering
% \includegraphics[width=1\textwidth]{figs/fig1.png}
% \caption{The Current State of Research in CRSs}
% \label{fig1}
% \end{figure*}

\begin{figure*}
    \centering
    \begin{subfigure}[b]{0.5\textwidth}
        \centering
        \includegraphics[width=\textwidth]{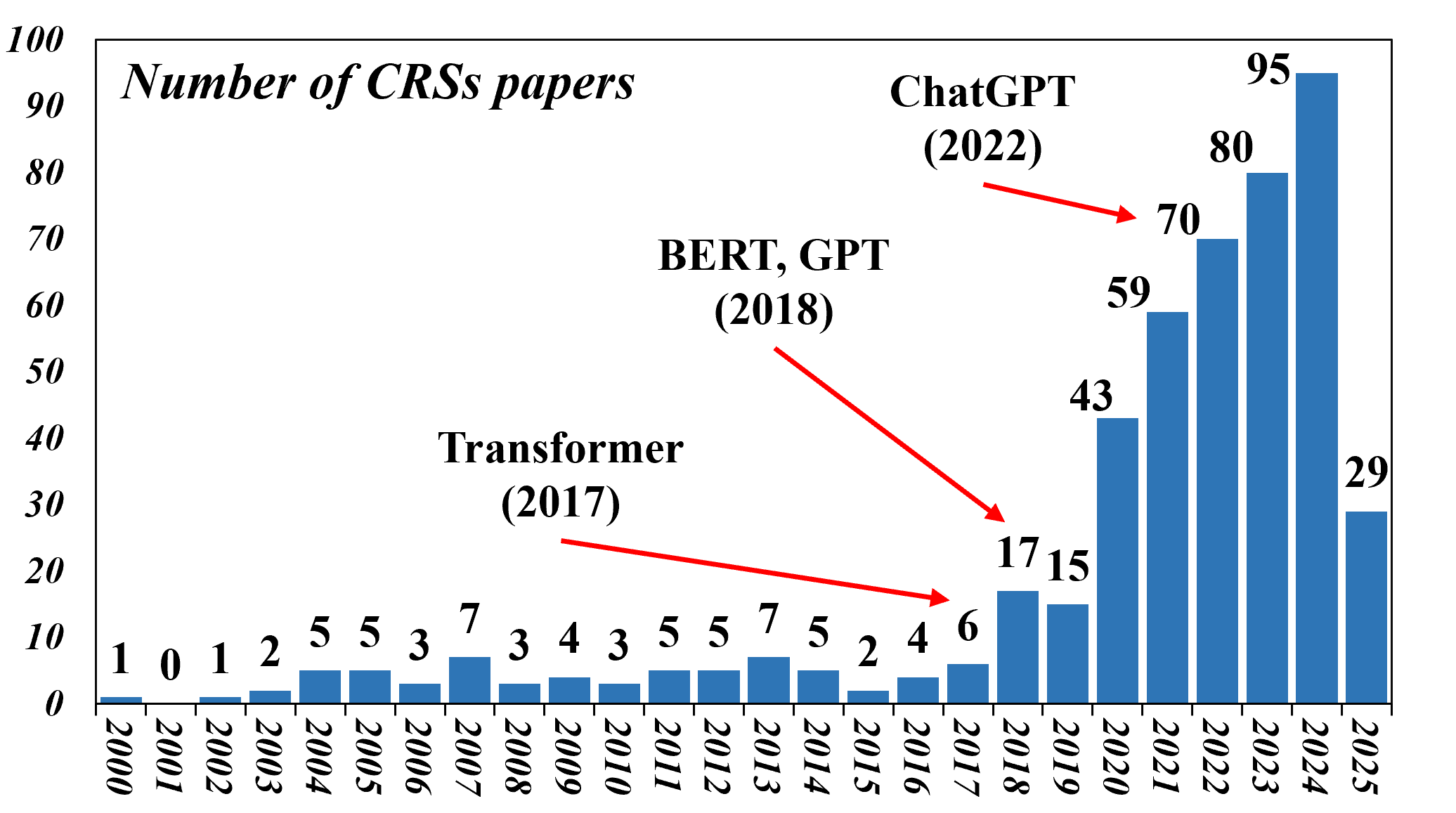}
        \caption{Development Trend and Key Time Nodes of CRSs.}
        \label{fig1a}
    \end{subfigure}
    ~~
    \begin{subfigure}[b]{0.48\textwidth}
        \centering
        \includegraphics[width=\textwidth]{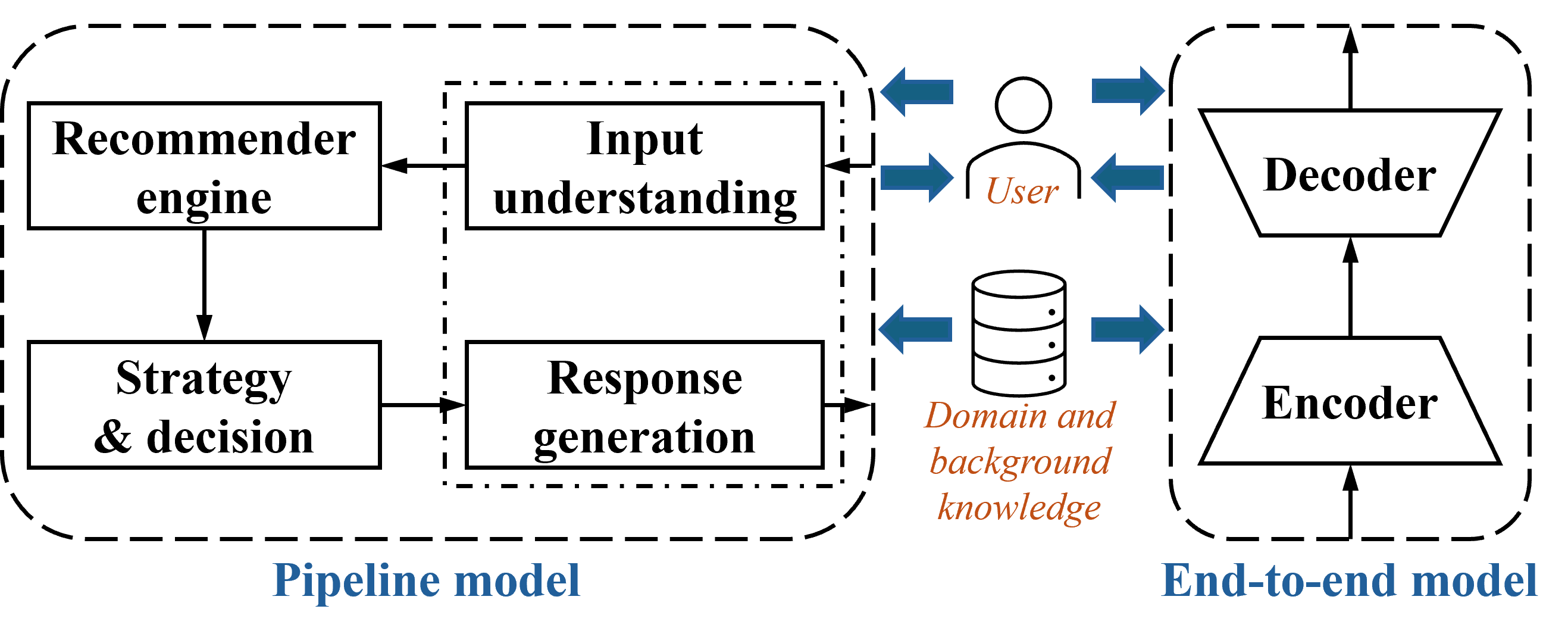}
        \caption{Technical Architecture of CRSs.}
        \label{fig1b}
    \end{subfigure}

    \caption{The Current State of Research in CRSs (The statistics of the publications are as of July 1, 2025).}
    \label{fig1}
\end{figure*}

\begin{table*}[htbp]
    \centering
    \caption{Simulation Methods in CRSs - Taxonomy Framework.}
    \label{tab1}
    % \footnotesize
    %\begin{tabular*}{0.99\linewidth}{@{}ccccc@{}}
    % \begin{tabular}{p{3.2cm}<{\centering} p{1.8cm}<{\centering} p{2.5cm}<{\centering} p{0.7cm}<{\centering} p{7cm}<{\raggedright}}
    \begin{tabular}{p{3.2cm}<{\centering} p{1.8cm}<{\centering} p{2.5cm}<{\centering} p{0.7cm}<{\centering} p{7cm}<{\centering}}
    
    \toprule
    %Topics & Publication &Source &Year &\multicolumn{1}{c}{Main contribution}\\
    Topics & Publication &Source &Year &Main contribution\\
    
    \midrule
    \multirow{2}{*}{\textit{Datasets Construction}} & \cite{liang-etal-2024-llm} & ACL &24 & A large-scale CRSs dataset constructed by leveraging LLMs.\\
    \cline{5-5}
    & \cite{kim-etal-2024-pearl} & ACL &24 & A Review-Driven Persona-Knowledge Grounded Conversational Recommendation Dataset.\\
    
    \cline{2-5}
    \multirow{2}{*}{\textit{Algorithm Design}} & \cite{wang-zhou-kdd23} & KDD &23 & Counterfactual data augmentation for CRSs. \\
    \cline{5-5}
    & \cite{shen-tors24} & ACM ToRS &24 & User simulator with fuzzy feedback.\\
   
    \cline{2-5}
    \multirow{8}{*}{\textit{System Evaluation}} & \cite{zhang-balog-kdd20} & KDD &20 & The first user simulation framework for CRSs evaluation, alternative of human evaluation.\\
    \cline{5-5}
    & \cite{zhangs-sigir22} & SIGIR &22 & Focus on simulating user utterances in CRSs.\\
    \cline{5-5}
    & \cite{Afzali-wsdm23} & WSDM &23 & An extensible user simulation toolkit.\\
    \cline{5-5}
    & \cite{wang-etal-2023-rethinking-evaluation} & EMNLP &23 & The first LLM-based user simulator for CRSs evaluation.\\
    \cline{5-5}
    & \cite{Bernard-cui24} & CUI &24 & Focus on robustness and effectiveness.\\
    \cline{5-5}
    & \cite{yoon-etal-2024-evaluating} & NAACL &24 & The first evaluation protocol for LLM-based user simulation in CRSs.\\
    \cline{5-5}
    & \cite{DiBratto2024} & UMUAI &24 & Use simulator to investigate the quality of the theoretical background.\\
    \cline{5-5}
    & \cite{zhu-wwwcomp25} & WWW &25 & Emphasize the Controllability, Scalability, Human-Involves of user simulation.\\
    
    \cline{2-5}
    \textit{Empirical Study} & \cite{zhu-wwwcomp24} & WWW &24 & Evaluate the evaluation methods for user simulators based on LLMs.\\
    
    \bottomrule
  %\end{tabular*}
  \end{tabular}
  
\end{table*}

\section{Method}

First, we conducted a title keyword search for ``conversation* recommend*'' across the dblp computer science bibliography. After removing preprints and duplicates, 447 relevant published articles on CRSs were retained. These articles cover key themes in CRSs research. Through iterative screening based on relevance criteria, 11 publications directly addressing simulation in CRSs were identified. To systematize findings, we developed a taxonomy framework aligned with core research objectives, organizing CRSs studies into four categories: (1) datasets construction, (2) algorithm design, (3) system evaluation, and (4) empirical studies. These 11 articles were mapped onto the framework to illustrate how simulation methods function across different research domains, as detailed in Table \ref{tab1}.

%  \verb+"+conversation* recommend*\verb+"+

\section{Findings}

\subsection{Simulation in CRSs Datasets Construction}

Data serves as a foundational component of intelligent systems research. While conventional RSs and dialogue systems benefit from established benchmarking datasets, CRSs specific datasets remain critically underdeveloped \cite{Jannach2023}. This methodological gap systematically constrains progress in both scholarly understanding and technical innovation within the domain. Not until 2018 was the first publicly available large-scale CRSs-specific dataset, REDIAL \cite{lir-nips18}, constructed. REDIAL collected data via Amazon Mechanical Turk (AMT) in a crowdsourced manner, comprising over 10,000 real world recommendation centered dialogues from 956 users. Prior to REDIAL, most CRSs research data was synthesized from conventional recommendation, everyday conversations, or shopping reviews datasets \cite{zhangyf-cikm18}. These synthetic datasets, either manually created or auto-generated \cite{Suglia-17}, were mostly rigid and inflexible. Although REDIAL's data, collected via crowdsourcing, was not gathered in real CRS settings, its real-human-generated dialogues and interactions provide greater flexibility and authenticity. Consequently, following REDIAL, more crowdsourced datasets have emerged, including topic guided and oriented multitype dialogs, covering various recommendation scenarios like movies and restaurants, among others \cite{GAO2021100}.

However, due to technological advancements and the growing data volume and quality requirements in CRSs research, datasets like REDIAL are no longer considered "large scale" and fail to meet the demands of data-driven system research, let alone real-world industrial applications. Most existing CRSs datasets suffer from data inextensibility and semantic inconsistency. Additionally, the cost of building datasets through crowdsourcing is rising sharply.

Generative models, particularly LLMs, exhibit superior capabilities in text understanding, dialogue generation, and role playing \cite{Shanahan2023}. Recently, some studies have employed LLM-based simulation methods to create conversational recommendation data, such as LLM-REDIAL \cite{liang-etal-2024-llm}. LLM-REDIAL generates multi-turn dialogues and recommendation feedback through instruction templates, assigning each dialogue turn a goal to ensure semantic consistency. The resulting synthetic dataset shows remarkable improvements in scale, text diversity, and dialogue quality. Moreover, LLM-based synthetic CRSs data offers significant cost advantages. For comparison, each conversation in REDIAL costs \$1, excluding data processing and labeling expenses. In contrast, LLM-REDIAL, comprising 47.6k multi-turn dialogues with 482.6k utterances across 4 domains, only costs \$750 in total.

\subsection{Simulation in CRSs Algorithm Design}

Studies in this category directly use simulation methods to address algorithmic issues in CRSs. For example, CFCRS enhances CRSs through counterfactual data augmentation \cite{wang-zhou-kdd23}. It leverages counterfactual learning to augment user preferences and uses the augmented preferences to simulate conversational data. CFCRS also performs interventions on the representations of target users' interacted entities and employs an adversarial training method with a curriculum schedule. This approach gradually optimizes the edit function, improving the recommendation capacity of CRSs. The User-centric User Simulator with Fuzzy Feedback (UUSFF) is a simulation framework that models user responses more naturally by incorporating fuzzy feedback based on users' inherent preferences \cite{shen-tors24}. This framework allows users to respond to attribute questions based on both the attribute-item relations of the target item and their long-term interests in related items. Integrated into the Multi-Interest Multi-Round conversational recommendation framework, UUSFF enhances the effectiveness of the recommendation algorithm.

\subsection{Simulation in CRSs Evaluation}

\subsubsection{Complexity challenges of CRSs Evaluation}

RSs are typically evaluated through three approaches: online evaluation (or human evaluation), offline evaluation, and user simulation. Online evaluation, considered the most ideal, is conducted in real-world settings using controlled experiments (A/B test) or quasi-experiments to assess algorithm performance and effectiveness. However, it faces significant challenges. The real environment is only accessible to a small number of researchers, and experiment deployment can negatively impact user experience, potentially eroding consumer trust, while also being costly and risky. Laboratory experiments, an alternative to online evaluation, use human subjects (not necessarily real users) without requiring a real environment, such as via AMT. Yet, this method faces issues with data quality control and still struggles with the scale-cost balance. Offline evaluation, the most common method in algorithm and intelligent system research, uses publicly available offline datasets and computational metrics without actual deployment or human participants. This simplicity, however, comes at the cost of reduced validity, especially for CRSs.

Unlike the mature evaluation protocols and methods for conventional RSs and conventional dialogue systems, no consensus has been reached for CRSs evaluation. This is due to the complexity of CRSs. Firstly, conversational recommendation incorporates more direct user experience elements \cite{jin-tors24}. Secondly, the interaction between users and the system is multi-turn and diverse. This complexity accentuates the high costs and ethical risks of online and laboratory experiments. It also exacerbates the limitations of single round evaluations and fixed ground truth in offline experiments.

\subsubsection{User simulation in CRSs evaluation}

In tackling the complexity challenges outlined in Section 3.3.1, a burgeoning area of research has centered on the development of user simulation methodologies for the evaluation of CRSs. User simulation leverages rule-based or algorithm-based user models to replicate authentic user interactions and feedback within offline settings. This approach offers significant advantages in terms of cost-effectiveness, flexibility, and adaptability, attributes that have propelled its growing prominence in the realm of information access system research \cite{Balog-zhai-f18}.

Krisztian Balog's team has made pioneering contributions to the evaluation of CRSs through user simulation. Their approach involves agenda-based simulation, which decomposes simulated users into four distinct modules: Natural Language Understanding (NLU), Natural Language Generation (NLG), preference model, and interaction model. The dialogue process is modeled as a Markov Decision Process (MDP), with a specifically designed CIR6 model for simulating the user's action space. This framework was initially applied to CRSs evaluation \cite{zhang-balog-kdd20}, and later evolved into UserSimCRS \cite{Afzali-wsdm23}, the first comprehensive simulation-based evaluation toolkit in this field. Expanding on their foundational work in user simulation, the researchers further investigated user utterance reformulation and CRSs breakdowns. They recognized that user simulators have limited robustness when the system fails to understand generated user utterances. To tackle this issue, they conducted a user study to capture real users' reformulation patterns and integrated these insights into a user simulator \cite{zhangs-sigir22}. They defined CRSs Breakdowns as system failures, unexpected or irrelevant replies, which can disrupt conversation flow. To enhance system robustness against such breakdowns, the team developed a diagnostic and development tool aimed at improving CRSs \cite{Bernard-cui24}.

The evaluation of CRSs primarily centers on human-computer interaction, unlike conventional RSs. Current user simulators have limitations in that they are static, restricted to fixed action sets, and lack generative capabilities. The advent of LLMs has opened up new possibilities for user simulation. iEvaLM represents the first approach to employ LLMs for evaluating CRSs \cite{wang-etal-2023-rethinking-evaluation}. By leveraging LLMs' remarkable instruction-following abilities, it devises user simulators, thereby liberating CRSs from rigid, human-crafted dialogue constraints and facilitating more natural and human-like interactions. Additionally, novel evaluation protocols have been put forward to gauge the accuracy of language models in mimicking human behavior within conversational recommendation settings (Yoon et al., 2024). These studies reveal that through careful prompting and model selection, improvements can be achieved in aspects such as item diversity, alignment with human preferences, request personalization, and feedback coherence. Furthermore, certain research endeavors have incorporated linguistic methods into CRS evaluation and developed LLM-based user simulators for assessment purposes \cite{DiBratto2024}, highlighting the multidisciplinary nature of CRSs evaluation.

\subsubsection{Simulation in CRSs Empirical Study}

User simulations can be utilized in experimental or empirical research for CRSs. For instance, given that simulation evaluation methodologies are still in a state of development, certain studies have directed their attention toward evaluating simulation methods based on LLMs. One such investigation, building upon the iEvaLM framework, executed an analytical validation study \cite{zhu-wwwcomp24}. It uncovered intrinsic limitations associated with employing LLMs to develop user simulators for CRSs and consequently put forward an enhanced solution.

Due to field experiments' constraints, such as cost, ethical risks, and counterfactuals, using simulated users for empirical research has long appealed to the RSs \cite{zhangjj-isr20}. The remarkable role-playing capabilities LLMs have revitalized the simulation of human behavior \cite{Shanahan2023}. Recently, there has been extensive research on LLM-based simulations across various domains, including economic society, information access systems (encompassing recommendation and search), and gaming \cite{Park-stanfordtown, wang-tois25, zhang-sigir24, zhanghr-www25}. This approach offers high fidelity, reproducibility, flexibility, low cost, and minimal ethical risks. Therefore, implementing LLM-based user simulations in CRSs environments could better explore general or specific empirical or experimental issues in recommendation. For instance, it could examine how emotions are transmitted between users and systems in CRSs.

\section{Challenges and Opportunities}

In the research on simulation for CRSs, challenges and opportunities are intertwined. For dataset construction and algorithm design focused on data simulation, the main challenge is the bias in seed datasets and its propagation during simulation. The advent of LLMs hasn't fully resolved this issue. In order to make the simulation system close to the real environment, multi-modal considerations need to be introduced in some application scenarios \cite{wang-muse-arxiv25}. Moreover, prompt templates in LLM-based simulations mostly depend on manual design, limiting output flexibility. Furthermore, a crucial challenge in algorithm design lies in the tight integration of simulation with recommendation processes. This integration should offer extra information and a correction basis for algorithm training while minimizing impact on efficiency, especially when LLMs are deployed.

In the research on system evaluation and empirical study in CRSs, user simulations, especially those based on LLMs, have certain advantages but still cannot directly replace real users. A common challenge is that LLM technology is still evolving. Most LLMs exhibit struggles with hallucination and causal reasoning. Existing LLM outputs are based on the text semantic space, and current simulations often neglect user behavior modeling. The gap between the text semantic space and behavioral semantics challenges the reliability of LLM-based behavior simulations. Although pre-trained LLMs are known for their world knowledge, most real users have a cognitive scope that cannot match this breadth. LLM-based user simulations often produce "cognitive supermen," leading to inaccurate evaluation and empirical research conclusions. Combining dialogue and recommendation areas makes it difficult to establish a universally accepted, unified evaluation protocol. Compared to current research that focuses on movie recommendations, simulations need to consider more diverse recommendation scenarios with unique logic and action spaces \cite{vlachou-arxiv24}. User simulations should also evolve from single-mode text-based interactions to multi-mode simulations for better alignment with real users.

\section{Conclusion}

We systematically examined simulation methods and their applications in CRSs. Our findings reveal that, within CRSs research, the significance of static offline data is gradually being supplanted by interactive simulation environments. As technology evolves, to meet the requirements of datasets construction, evaluation protocols, and empirical research in CRSs, researchers are striving to strike a balance between costly yet authentic human participation and limited but simple offline data. Presently, despite numerous controversies and challenges, simulation seems to be the most promising path forward.

\begin{acks}
This study is supported by the National Natural Science Foundation of China (No.72171165).
\end{acks}

\bibliographystyle{SageV}
\bibliography{ref}

\end{document}